# Generation of correlated photon pairs in a chalcogenide $As_2S_3$ waveguide

C. Xiong[1, 2], G. D. Marshall[1, 3], A. Peruzzo[4], M. Lobino[4], A. S. Clark[4], D.-Y. Choi[1, 5], S. J. Madden[1, 5], C. M. Natarajan[6], M. G. Tanner[6], R. H. Hadfield[6], S. N. Dorenbos[7], T. Zijlstra[7], V. Zwiller[7], M. G. Thompson[4], J. G. Rarity[4], M. J. Steel[1, 3], B. Luther-Davies[1, 5], B. J. Eggleton[1, 2] and J. L. O'Brien[4]

[1]*Centre for Ultrahigh-bandwidth Devices for Optical Systems (CUDOS), Australia*
[2]*Institute for Photonics and Optical Science (IPOS), School of Physics, University of Sydney, New South Wales 2006, Australia*
[3]*MQ Photonics, Dept. of Physics & Astronomy, Macquarie University, NSW 2109 Australia*
[4]*Centre for Quantum Photonics, H. H. Wills Physics Laboratory*
*Dept. of Electrical and Electronic Engineering, University of Bristol, Bristol BS8 1UB, UK*
[5]*Laser Physics Centre, Australian National University, Canberra ACT 0200, Australia*
[6]*School of Engineering and Physical Sciences, Heriot-Watt University, Edinburgh EH14 4AS, UK*
[7]*Kavli Institute for Nanoscience, Delft University of Technology, Lorentzweg 1, 2628 CJ Delft, the Netherlands*

chunle@physics.usyd.edu.au

We demonstrate the first 1550 nm correlated photon-pair source in an integrated glass platform—a chalcogenide $As_2S_3$ waveguide. A measured pair coincidence rate of 80 per second was achieved using 57 mW of continuous-wave pump. The coincidence to accidental ratio was shown to be limited by spontaneous Raman scattering effects that are expected to be mitigated by using a pulsed pump source.

Key to emerging quantum photonic technologies such as quantum communication[1] and quantum computation[2] is the availability of bright and low-noise integrated photon-pair sources in the 1550 nm wavelength band. The generation of 1550 nm correlated photon pairs from photonic-chip platforms, based on parametric down conversion in periodically poled lithium niobate (PPLN) waveguides[3] and spontaneous four-wave mixing (SFWM) in silicon nanowires[4–6], has been widely studied. The PPLN platform requires an excess temperature control unit to achieve phase matching and thus is incompatible with integration. The silicon platform is promising because of its mature fabrication technology; however, the two-photon absorption (TPA) induced long life-time free carriers in silicon result in free-carrier absorption (FCA), which may degrade the source quality[4]. Thus there remains a need for 1550 nm correlated photon-pair generation from a platform that supports integration and has no free carriers.

Recent work suggests that the chalcogenide $As_2S_3$ platform has several potential advantages for integrated 1550 nm correlated photon-pair sources[7]. Firstly, $As_2S_3$ glass exhibits high third-order nonlinearity (comparable to Si)[8] and four-wave mixing experiments in low-loss $As_2S_3$ waveguides have demonstrated high parametric gain[9]. Secondly, this glass has negligible TPA making it immune to FCA[8]. Furthermore, $As_2S_3$ glass has a low Raman-gain window, which provides a potential means to reduce spontaneous Raman scattering (SpRS) induced noise[7]. Finally, the glass has the potential to be made into more complex photonic circuits consisting of wavelength division devices and 2D/3D structures. In this paper, we report the first observation of 1550 nm correlated photon-pair generation by SFWM in an integrated chalcogenide $As_2S_3$ waveguide using a continuous-wave (CW) pump. By characterizing the dependence of pair coincidence rate on pump power and waveguide output coupling efficiency, we show that the measured net pair coincidence rate of 80 $s^{-1}$ is comparable with that reported in silicon[4, 5]. The coincidence to accidental ratio (CAR), defined as the ratio between the net coincidence rate and the accidentals rate, was found to be limited by (SpRS), but we expect improvement by using a pulsed pump source so that the pair production is time gated.

Figure 1 shows the experimental setup. The waveguide is 7.1 cm long. Both facets of the waveguide were infrared anti-reflection coated to improve coupling and reduce Fabry-Pérot resonances. Lensed fibers with a mode field diameter of 2.5 μm were used to couple to the waveguide. At 1550 nm the total insertion losses, including propagation and coupling losses, for the fundamental TE and TM modes were measured to be 14.24 and 18.6 dB, respectively. The propagation losses of the TE and TM modes were estimated to be 0.7 and 1.3 dB $cm^{-1}$. The dispersion at 1550 nm for the TE and TM modes was estimated by a commercial finite-element package (*RSoft FemSIM*) to be -239 and 22 ps $nm^{-1}km^{-1}$, respectively. The nonlinear coefficient was estimated as $\gamma$=14 $W^{-1}m^{-1}$, based on the nonlinear refractive index $n_2$=3×$10^{-18}$ $m^2W^{-1}$ and an effective mode area of 0.86 $\mu m^2$.

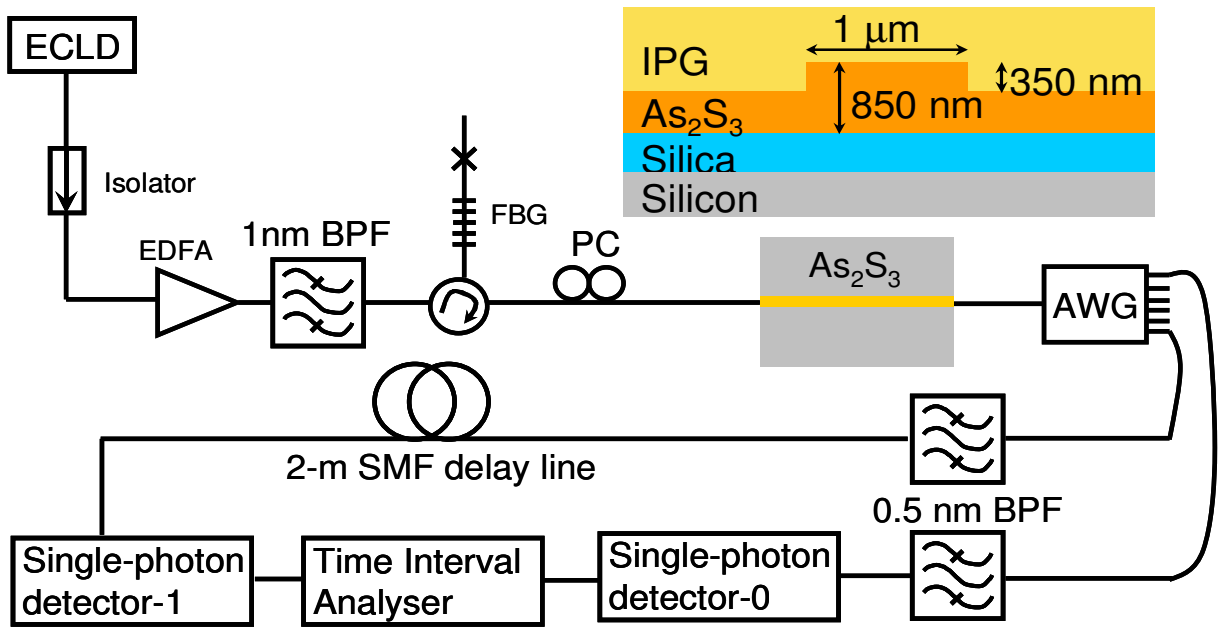

FIG. 1. Experimental setup. ECLD: external-cavity laser diode, EDFA: erbium-doped fiber amplifier. A band-pass filter (BPF) with a full width at half maximum (FWHM) of 1 nm, a circulator and a fiber Bragg grating (FBG, reflection FWHM 0.5 nm) were employed to reduce noise from the ECLD and EDFA. PC: polarization controller, AWG: arrayed waveguide grating. Inset shows the cross-section of the $As_2S_3$ waveguide. IPG: inorganic polymer glass, SMF: single-mode fiber.

With these parameters and an in-waveguide power of approximately 60 mW (see below), we expect generation of correlated photon pairs by SFWM spaced equally in frequency about the pump. We refer to the higher (lower) frequency photon as the signal (idler). The output light from the $As_2S_3$ chip was fed into an arrayed waveguide grating (AWG), which separated pump and pair photons. The AWG had 40 channels with frequency spacing 100 GHz and channel FWHM 50 GHz. A tunable BPF with a FWHM of 0.5 nm was inserted into each of the expected pair photon channels to further reduce the out-of-band noise. The losses of the AWG and BPF were measured to be 6.51 and 6.75 dB for the two pair-photon channels. After one of the pair photons was optically delayed, they were detected by two fiber-coupled superconducting single-photon detectors (SSPDs) based on NbTiN nanowires[10, 11]. The measured detection efficiencies (at 1550 nm wavelength and 1000 $s^{-1}$ dark count rate) were 18% for detector-0 and 8% for detector-1. The output signals from SSPDs were used as start and stop pulses for a time interval analyzer (TIA).

To produce pairs, we fixed the pump wavelength at 1549.315 nm, corresponding to the center of one AWG channel. We controlled the signal frequency shift $\nu$ from the pump channel using the AWG and BPF with a maximum detuning of $\nu=\pm1.4$ THz. All of our experiments used TE polarized pump light as this mode had the lowest propagation loss and provided a SFWM bandwidth that covered the whole of the AWG frequency range. Figure 2 shows a typical time-bin histogram for coincidence measurements when the waveguide-coupled power was $P_0$=57 mW (assuming equal division of the coupling losses between input and output) and $\nu$=1.4 THz. The time bin was 16 ps and the data collection time was 300 s. The sharp peak, located at a delay of 11.1 ns corresponding to the 2 m SMF delay line, clearly indicates the generation of correlated photon pairs. The noise floor is due to the accidental events recorded by the TIA and the FWHM of the peak was approximately 200 ps. The idler and signal counting rates at the two detectors were $N_0=3.45\times10^6$ $s^{-1}$ and $N_1= 1.34\times10^6$ $s^{-1}$. We analyze the data in terms of the following equations[7, 12]:

$$
\begin{aligned}
r &= \Delta\nu(\gamma P_0 L_{\mathrm{eff}})^2 \mathrm{sinc}^2[\beta_2(2\pi\nu)^2 L/2 + \gamma P_0 L], \\
C &= \sigma\eta_\alpha^2\eta^2\eta_0\eta_1 r, \\
N_0 &= \sigma\eta\eta_0(\eta_\alpha r + r_{n0}) + d_0, \\
N_1 &= \sigma\eta\eta_1(\eta_\alpha r + r_{n1}) + d_1, \\
A &= N_0 N_1 t, \\
\mathrm{CAR} &= C/A,
\end{aligned}
\qquad (1)
$$

where $r$ is the rate of photon pair generation by SFWM, $\Delta\nu$ is the bandwidth of the pair-photon channel, $L$ is the waveguide length and $L_{\mathrm{eff}}=[1-\exp(-\alpha L)]/\alpha$ is the effective waveguide length accounting for the propagation loss $\alpha$, $\beta_2=3.048\times10^{-25}$ $s^2m^{-1}$ is the second-order dispersion parameter, $C$ is the net coincidence rate, $\eta_\alpha$ is a parameter taking into account the pair-photon loss in the waveguide and can be determined by comparing the measured coincidence rate $C$ and the calculated photon pair generation rate $r$, $\eta$ is the waveguide output coupling efficiency, $\eta_0$=0.04 and $\eta_1$=0.017 are the photon collection efficiencies, including the AWG, BPF transmission efficiency and the SSPDs quantum efficiency, $r_{\mathrm{ni}}$ (i=0, 1) is the rate of noise-photon generation, $d_0=d_1=1000$ $s^{-1}$ are the rate of SSPDs dark count, $A$ is the accidentals rate in the time-bin duration of $t$ and $\sigma$ is a duty cycle parameter which for CW operation is 1. Finally CAR is the coincidence to accidentals ratio introduced earlier.

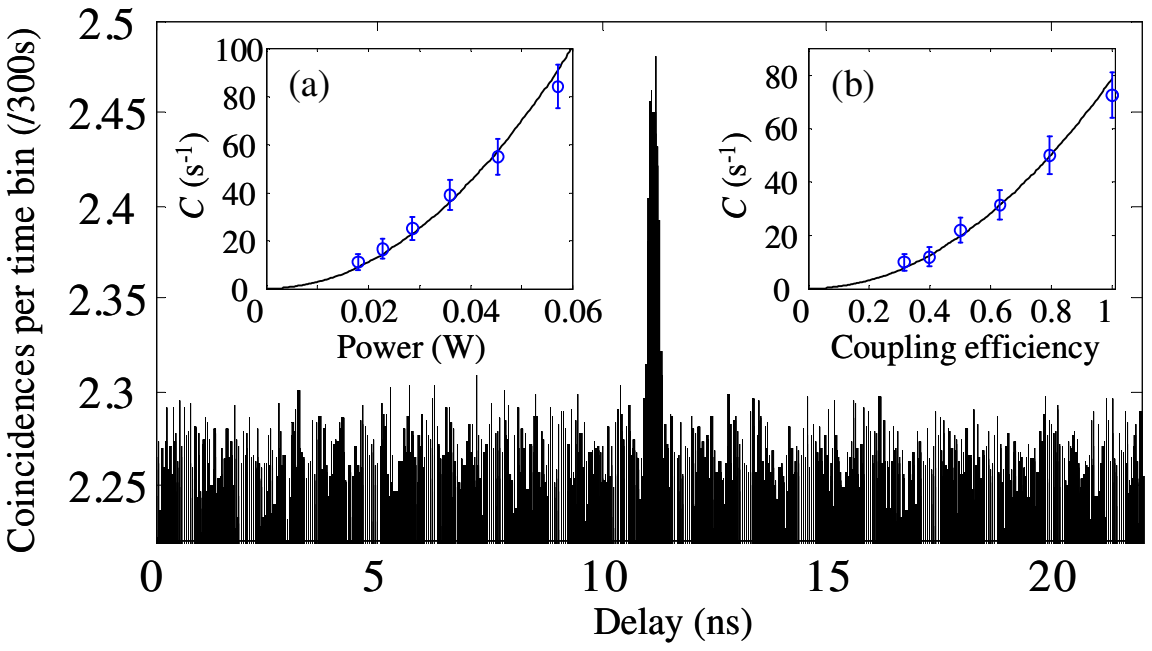

FIG. 2. A typical time-bin histogram for coincidence measurements. Insets show the measured net coincidence rate as a function of (a) pump power and (b) normalized output coupling efficiency (normalized to the optimum). Circles are the measured data. Poisson error bars are the square root of the data. The solid lines are quadratic fits following Eqs. (1), for the coupled pump power and normalized output coupling efficiency, respectively.

Equations (1) indicate that the net pair coincidence rate should depend quadratically on the coupled pump power and the output coupling efficiency. To confirm this we investigated these dependencies for $\nu$=1.4 THz. In the first case (Fig. 2(a)), we varied the pump power and kept all other conditions unchanged. In the second case (Fig. 2(b)), we varied the output coupling by misaligning the waveguide and the lensed fiber. At each point we obtained a time-bin histogram and extracted the net coincidence rate by subtracting the accidentals rate from the coincidence peak. The data presented in the insets of Fig. 2 confirm the expected quadratic dependencies.

Pair brightness and correlation are the two standard measurements of the quality of a photon-pair source. The maximum measured pair coincidence rate was about 80 $s^{-1}$. This is comparable with that reported in silicon[4, 5], and could be further improved with enhanced waveguide coupling enabled by inverse taper coupling sections[4–6]. The pair correlation is experimentally characterized by the CAR. As can be seen from Fig. 2, the coincidence peak is about 6% above the noise floor, which indicates a low CAR due to the high level of noise. The possible noise sources are the residual pump photons, photons generated in the silica-fiber connecting components, and SpRS in the chalcogenide waveguide itself. The first two possibilities were easily eliminated through tests by using one additional pump rejection FBG, and removing the chalcogenide waveguide from the setup, respectively. To determine if SpRS in the waveguide itself was the dominant noise, we measured the singles count rates as a function of pump power at $\nu$=1.4 THz. Figure 3 shows the measured and predicted singles count rates due to SFWM (solid lines) and SpRS (dotted lines) at different pump power[7]. It is clear that most of the singles count rate is attributable to SpRS. The count rate difference between signal and idler channels was due to the different quantum efficiency of the two detectors and was confirmed by repeating the measurement with the detector positions swapped. It is important to note that the SpRS contribution to the singles count rates is almost the same for both the signal and idler channels, a phenomenon expected when the frequency detuning $\nu$ is small[12]. We also investigated the dependence of CAR on the frequency detuning $\nu$. We found that the CAR was relatively constant over the range of 0.6–1.4 THz. Below $\nu$=0.6 THz, the CAR dropped quickly as the impact of residual pump photons transmitted by the AWG and BPF becomes significant close to the pump wavelength.

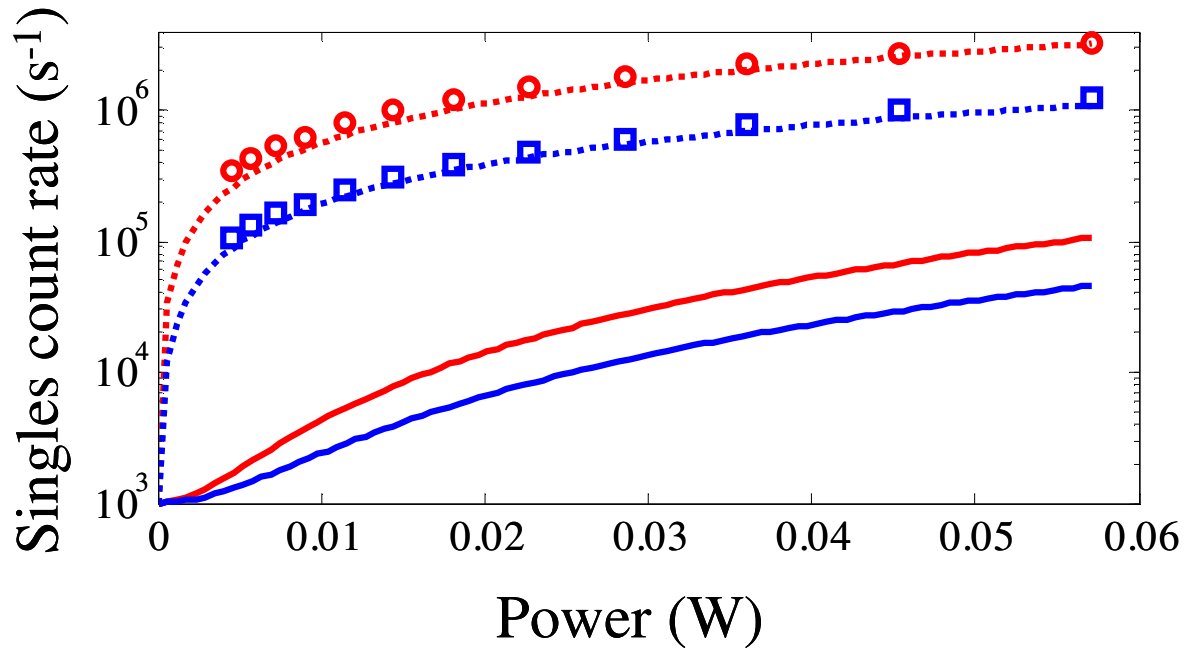

FIG. 3. Singles count rate as a function of pump power. Circles and squares show the measured idler and signal count rates. Lines show the predicted count rates for idler (red) and signal (blue) using Eqs. (1) and SpRS equations in Ref. 7.

Figure 3 indicates that the noise performance of our source was principally limited by the generation of SpRS noise, a phenomenon expected in glass platforms with a CW pump[13]. In the future, the use of a pulsed pump laser will provide time-gated pair production and immediately improve the CAR. This can be understood by considering Eqs. (1) in a pulsed pump regime for which the duty cycle parameter becomes $\sigma=\tau B$, where $\tau$ is the pulse duration and $B$ is the repetition rate. Observe that the net coincidence rate $C$ depends on $\sigma$ linearly, but the accidentals rate $A$ depends on $\sigma^2$, thus intuitively the CAR is proportional to $1/\sigma$. To illustrate this, we plot the predicted CAR as a function of average number pairs per pulse in Fig. 4 (solid line). The calculation is based on Eqs. (1) and all parameters used in our experiment, with $\tau$=5 ps and $B$=100 MHz. It can be seen that the CAR can be as high as 50 at an average pair generation rate of 0.01 per pulse. Our previous theoretical analysis has shown that the $As_2S_3$ glass has a low Raman gain window at a shift of approximately $\nu$=7.4 THz for which the CAR is predicted to be much improved[7]. This experiment, however, did not operate in this regime due to the limited SFWM bandwidth of our existing waveguide device. In future experiments, we plan to lower the dispersion of the waveguide at 1550 nm by modifying its cross section size to extend the SFWM bandwidth so that we can produce photons in the 7.4 THz low Raman noise window[7]. The dotted line in Fig. 4 predicts the CAR in such a dispersion engineered waveguide. The CAR is expected to be 250 at a level of 0.01 pairs per pulse. Chalcogenide waveguides are particularly attractive in the pulsed regime, as the lack of FCA removes a limit on the peak pump power which is then determined only by the requirement to keep double pair production at an acceptable level.

In conclusion, we have demonstrated correlated photon-pair generation in the 1550 nm wavelength band from an integrated chalcogenide $As_2S_3$ glass waveguide. The measured pair coincidence rate of 80 $s^{-1}$ is comparable with results in silicon waveguides. The CAR was limited by the SpRS noise in the CW pump scheme. Using a pulsed pump can immediately enhance the CAR to 50 at an average pair generation rate of 0.01 per pulse. Superior waveguide design including dispersion engineering and inverse taper coupling should provide additional improvement. Combined with its suitability for compact nonlinear processing components[8, 9], we believe that this work opens the way for $As_2S_3$ and related nonlinear glasses to play a significant role in quantum information processing.

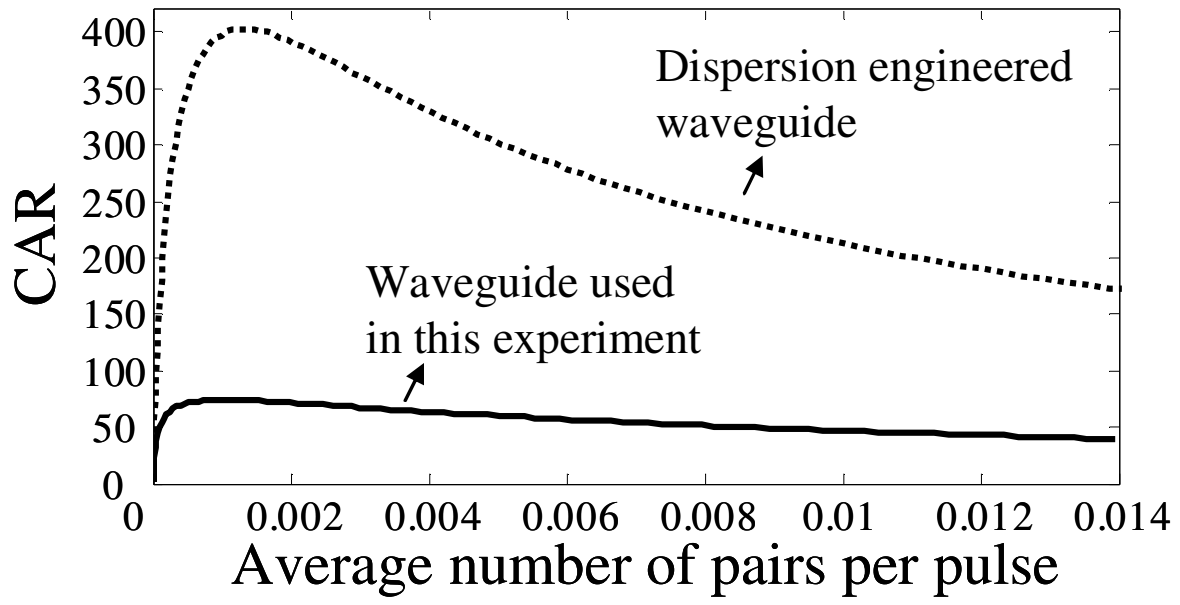

FIG. 4. The predicted CAR as a function of average number of pairs per pulse in a pulsed pump scheme.

We acknowledge the support of the Australian Research Council Centre of Excellence and Federation Fellowship programs; the Australian Academy of Science's International Science Linkages scheme; the Royal Society of London University Research Fellowship [RHH]; VIDI funding by FOM [SND, TZ and VZ]; and the Bristol Centre for Nanoscience and Quantum Information. CMN, MGT and RHH acknowledge support from EPSRC (UK) and assistance in constructing the SSPDs from Dr S. Nam at NIST, USA.

[1] N. Gisin, and R. Thew, Nat. Photonics **1**, 165 (2007).
[2] J. L. O'Brien, A. Furusawa and J. Vučković, Nat. Photonics **3**, 687 (2009).
[3] Myrtille Hunault, Hiroki Takesue, Osamu Tadanaga, Yoshiki Nishida, and Masaki Asobe, Opt. Lett. **35**, 1239 (2010).
[4] Jay E. Sharping, Kim Fook Lee, Mark A. Foster, Amy C. Turner, Bradley S. Schmidt, Michal Lipson, Alexander L. Gaeta, and Prem Kumar, Opt. Express **14**, 12388 (2006).
[5] S. Clemmen, A. Perret, S. K. Selvaraja, W. Bogaerts, D. van Thourhout, R. Baets, Ph. Emplit, and S. Massar, Opt. Lett. **35**, 3483 (2010).
[6] Ken-ichi Harada, Hiroki Takesue, Hiroshi Fukuda, Tai Tsuchizawa, Toshifumi Watanabe, Koji Yamada, Yasuhiro Tokura, and Sei-ichi Itabashi, IEEE J. Sel. Top. Quantum Electron. **16**, 325 (2010).
[7] C. Xiong, L. G. Helt, A. C. Judge, G. D. Marshall, M. J. Steel, J. E. Sipe, and B. J. Eggleton, Opt. Express **18**, 16206 (2010).
[8] V. G. Ta'eed, N. J. Baker, Libin Fu, K. Finsterbusch, M. R.E. Lamont, D. J. Moss, H. C. Nguyen, B. J. Eggleton, D. Y. Choi, S. Madden, and B. Luther-Davies, Opt. Express **15**, 9205 (2007).
[9] M. R. E. Lamont, B. Luther-Davies, D.Y. Choi, S. Madden, X. Gai and B. J. Eggleton, Opt. Express **16**, 20374 (2008).
[10] M. G. Tanner, C. M. Natarajan, V. K. Pottapenjara, J. A. O'Connor, R. J. Warburton, R. H. Hadfield, B. Baek, S. Nam, S. N. Dorenbos, E. Bermúdez Ureña, T. Zijlstra, T. M. Klapwijk, and V. Zwiller, Appl. Phys. Lett. **96**, 221109 (2010).
[11] S. N. Dorenbos, E. M. Reiger, U. Perinetti, V. Zwiller, T. Zijlstra, and T. M. Klapwijk, Appl. Phys. Lett. **93**, 131101 (2008).
[12] Q. Lin, F. Yaman, and Govind P. Agrawal, Phys. Rev. A **75**, 023803 (2007).
[13] J. G. Rarity, J. Fulconis, J. Duligall, W. J. Wadsworth and P. St. J. Russell, Opt. Express **13**, 534, (2005).